\documentclass[sigconf]{acmart}
\usepackage{tabularx}
\usepackage{adjustbox}
\usepackage{booktabs} 
\usepackage{comment}
\usepackage[framemethod=tikz]{mdframed}
\usepackage{enumerate}
\usepackage{pgfplots}

\usepackage[colorinlistoftodos]{todonotes}

\setcopyright{none}

\acmConference[MET 2018]{International Workshop on Metamorphic Testing}{May 2018}{Gothenburg, Sweden} 
\acmYear{2018}
\copyrightyear{2018}

\begin{document}
\title[Metamorphic Testing of a Biomedical Text Processing Tool]{Quality Assurance of Bioinformatics Software: A Case Study of Testing a Biomedical Text Processing Tool Using Metamorphic Testing}
\titlenote{Produces the permission block, and
  copyright information. 
}

\author{Madhusudan Srinivasan}
\authornote{The authors wish it to be known that, in their opinion, the first two authors should be regarded as joint First Authors.}
\affiliation{%
  \institution{Montana State University}
  \city{Bozeman}
  \state{Montana}
  }
\email{madhusuda.srinivasan@montana.edu}

\author{Morteza Pourreza Shahri}
\affiliation{%
  \institution{Montana State University}
  \city{Bozeman}
  \state{Montana}
 }
 \email{mpourrezashahri@montana.edu}

\author{Indika Kahanda}
\affiliation{%
  \institution{Montana State University}
    \city{Bozeman}
  \state{Montana}
 }
 \email{indika.kahanda@montana.edu}

\author{Upulee Kanewala}
\authornote{Corresponding author}
\affiliation{%
 \institution{Montana State University}
 \city{Bozeman}
 \state{Montana}
}
\email{upulee.kanewala@montana.edu}

\begin{abstract}
Bioinformatics software plays a very important role in making critical decisions within many areas including medicine and health care. However, most of the research is directed towards developing tools, and little time and effort is spent on testing the software to assure its quality. In testing, a test oracle is used to determine whether a test is passed or failed during testing, and unfortunately, for much of bioinformatics software, the exact expected outcomes are not well defined. Thus, the main challenge associated with conducting systematic testing on bioinformatics software is the oracle problem.

Metamorphic testing (MT) is a technique used to test programs that face the oracle problem. MT uses metamorphic relations (MRs) to determine whether a test has passed or failed and specifies how the output should change according to a specific change made to the input. In this work, we use MT to test LingPipe, a tool for processing text using computational linguistics, often used in bioinformatics for bio-entity recognition from biomedical literature. 

First, we identify a set of MRs for testing any bio-entity recognition program. Then we develop a set of test cases that can be used to test LingPipe's bio-entity recognition functionality using these MRs. To evaluate the effectiveness of this testing process, we automatically generate a set of faulty versions of LingPipe. According to our  analysis of the experimental results, we observe that our MRs can detect the majority of these faulty versions, which shows the utility of this testing technique for quality assurance of bioinformatics software.
\end{abstract}

%
%

\keywords{Software testing, Metamorphic testing, Bioinformatics, Biomedical Natural Language Processing}

\renewcommand{\shortauthors}{Srinivasan et al.}
\maketitle

\section{Introduction}
Researchers studying living organisms often rely on bioinformatics tools to answer research questions and develop conclusions~\cite{Bayat}. These bioinformatics tools help to create considerable biomedical research breakthroughs involving genomes~\cite{surpreet}. For example, bioinformatics tools are used in predicting the functions and modeling the structure of proteins~\cite{Bayat} as well as drug discovery~\cite{surpreet}. At the same time, lab experiments and clinical investigations  are highly time consuming and may require up to 15 years and millions to billions of dollars for introducing a compound to the market~\cite{surpreet}. Hence bioinformatics tools can help in analyzing the data faster and eventually reduce the time for discovery. This as a result can reduce the cost for researchers as well as the industry.

The use of software systems have grown exponentially over the years~\cite{Wrightbug}. But, the number of problems due to software faults have also increased in recent times~\cite{Wrightbug}. Thus, software testing plays a major role in detecting as many software faults as possible~\cite{10.1109/TSE.2010.62}. Bioinformatics programs are complex and require novel techniques to test them~\cite{kamali2015test}. Typically, these tools are developed by researchers for the analysis of biological and biomedical data but they are not tested adequately with varied and all possible levels of test data either due to time or budget constraints~\cite{Sanders}. Previous studies have shown that bioinformatics tools are not rigorously tested before being delivered to the end users since it is hard to validate the correctness of the output~\cite{Altschul}. Furthermore, testing of bioinformatics tools may require extensive interactions between the testers and the domain experts. The lack of communication can lead to the failure of using unit testing for important modules in the tool which reduces the reliability of the bioinformatics tools~\cite{David}.

Test oracle is an essential part of software testing and it is a mechanism for verifying the correctness of test output for a given set of test inputs. Many bioinformatics programs fall into the category of non-testable programs, where either a test oracle is unavailable or it is practically difficult to develop test oracles~\cite{Weyuker01111982}. Metamorphic testing (MT) is a testing technique that can alleviate the test oracle problem~\cite{MTRewiewTY2017}. Metamorphic testing involves identifying a set of properties from the program under test known as metamorphic relations (MRs).  Multiple test cases are executed and results of the test cases are used to check whether they satisfy or violate the identified MR. Violation of MRs indicate faults in the program. We hypothesized that MT can be an effective approach for testing bioinformatics tools.

In this study, we explore the feasibility and effectiveness of using MT for bioinformatics software. In particular, we identify metamorphic relations for the task of biomedical natural language processing. We use MT to test LingPipe~\cite{lingpipe}, a tool for processing text using computational linguistics and often used in bioinformatics for bio-entity recognition. Bio-entity recognition is the process of extracting biomedical terms from text, and assigning them to appropriate categories~\cite{lingpipe}. Bio-entity recognition tools such as LingPipe use named-entity detection to extract bio-entities such as genes, organisms, and diseases from very large corpora of biomedical literature~\cite{lingpipe}. The typical outputs from a bio-entity recognition process are hard for developers to validate and requires the support from domain experts. 
The effectiveness of our proposed metamorphic relations are evaluated using mutants~\cite{10.1109/TSE.2010.62}. Our result show that proposed MRs identify a majority of the mutants used in the evaluation and indicates that MT is an effective method for testing biomedical natural language processing software. 

The rest of this paper is organized as follows. Section \ref{sec:background} describes the background related to test oracle, metamorphic testing and bio-entity recognition. Section \ref{section_mt} provides details of each of the MRs we identify for testing bio-entity recognition software. Section \ref{sec:methodology} provides the methodology for validating the LingPipe tool using the proposed metamorphic relations. Sections \ref{sec:expsetup} and \ref{sec:results} provides the experimental setup and the results, respectively. Section \ref{sec:discussion} discusses the key observations from experimental results while Section \ref{sec:relatedworks} provides related works. Section \ref{sec:conclusions} concludes the paper and discusses promising future directions.

\section{Background}
\label{sec:background}

In this section, we provide some background on test oracles, metamorphic testing, and bio-entity recognition. The test oracles section focuses on the definition of an oracle, test oracle problem and test oracle problem related to our case study tool. The metamorphic testing section describes the step by step approach for conducting metamorphic testing. The Bio-entity recognition section provides a brief description of Named Entity Recognition and its utilization in the biomedical area.  

\subsection{Test Oracles}
\emph{Test oracle} is a mechanism used to verify the correctness of the output produced by a program~\cite{Weyuker01111982}. The \emph{Oracle problem} occurs when there is not an existing oracle or when it is practically difficult to determine the correct output ~\cite{Weyuker01111982}. The oracle problem is a common problem in testing scientific software~\cite{kanewala2014testing}. Programs which face the oracle problem are called \emph{non-testable programs}. Scientific softwares such as LingPipe used in our case study fall into the category of non-testable softwares. The LingPipe outputs are long and complex which makes it difficult to clearly define a test oracle. 

\subsection{Metamorphic Testing}
Suppose that a function $f$ is implemented by a program $p$. Steps for conducting metamorphic testing on $p$ is as follows:

 \begin{enumerate}
\item Identify MRs for function $f$.
\item Create source test cases. Source test cases $t_{s}$ is derived using the traditional test case selection methodology. Most common method for test case selection is random testing.   
\item Follow-up test case $I_{2}$ is derived from $I_{1}$  by applying MR to the source test case  $I_{1}$.
\item Execute test case $I_{1}$ and $I_{2}$ and get output $O_{1}$ and  $O_{2}$.
\item If $O_{1}$ and $O_{2}$ results does not satisfy the MR, then a fault is assumed to be present in the program.
\end{enumerate}
Consider a function that calculates the sum of integer elements in an array. An MR for this summation function can be defined using the property that shuffling the array elements should not change the actual sum of those elements. In order to test this function using this MR, an array containing some integers is used as a source test case. The follow up test case is a shuffled version of that same array. The outputs (in this case sums) obtained from the execution of source and follow up test cases is compared. MR violation occurs when a difference between the outputs are observed.

\subsection{Bio-entity Recognition}
Bio-entity recognition is the task of identifying and/or extracting biomedically important entities such as gene and protein names from text. LingPipe is a natural language processing tool often used for the bio-entity recognition by the bioinformatics community. LingPipe consists of a variety of modules which perform different tasks such as finding the names of people, organizations, or biomedical terms~\cite{lingpipe}. 
One of the main modules of LingPipe is the "Named Entity Recognition (NER)" module. The NER module performs the supervised training of a model, often a statistical model, and extracts the terms mentioned using the trained model~\cite{lingpipe}. Since this model can be trained on biomedical data, LingPipe is also able to extract bio-entities. The following example shows the protein name extracted from the given sample text using LingPipe.
\\
\begin{mdframed}[hidealllines=true,backgroundcolor=gray!20]
\textbf{Sample Sentence (PMID: 29320757):}\\``Neuritin plays an important role in the development and regeneration of the nervous system, and shows good prospects in the treatment and protection of the nervous system''.
\\
\\
\textbf{LingPipe output:}\\
 Term = Neuritin, Position = (0, 8)      
\end{mdframed}

As shown in this example, LingPipe returns the occurrences of bio-entities within the input text. The returning data contains a list of terms and corresponding positions. The term can be a single word or a combination of multiple words. The position indicates the start and end indices of the corresponding terms occurring within the input text. 
When applied to a large corpus of text, LingPipe returns a very large number of bio-entities and introduces difficulty in validating the results. Hence, the oracle problem persists with bio-entity recognition tools such as LingPipe. 

\section{METAMORPHIC RELATIONS FOR BIO-ENTITY RECOGNITION}
\label{section_mt}
We propose three categories of MRs for testing bio-entity recognition software. These categories are Addition, Deletion, and Shuffling relations. In the Addition relations, we usually extend a span of text by adding it to another span. For example, a sentence can be appended to another sentence, a sentence to a paragraph, a paragraph to an article, or a list of random words to another list of random words. In the Deletion relation, a span of text is truncated by removing a part of it. For instance, removing a consecutive list of words from a sentence, removing a sentence from a paragraph, removing a paragraph from an article, or removing a part from a list of random words. In the Shuffling relations, a span of text gets a new form. Rearranging the paragraphs of an article, or all the words of a span of text are examples of Shuffling relations. 

In the following definitions, each extracted biomedical entity~(BE) contains a term and the position of the extracted biomedical entity in the original input text, which are referred to as $BE_t$ and $BE_p$, respectively. Further, $length(T)$ indicates the length of the text $T$ in number of charters.

\subsection{Addition Relations}

\subsubsection{MR1: Adding a sentence to another sentence}
Given two sentences $S_1$ and $S_2$, we append $S_2$ to $S_1$. 
Referring to the new text as $S\textprime$,
\begin{equation*}
\label{eq:MR1t}
BE_t(S\textprime) = BE_t(S_1) \cup BE_t(S_2)
\end{equation*}
\begin{multline*}
\label{eq:MR1p}
BE_p(S\textprime) = \{x \mid x\ \in BE_p(S_1)\} \cup \{x+length(S_1) \mid  x \in BE_p(S_2) \}
\end{multline*}


\subsubsection{MR2: Adding a sentence to a paragraph}
Given a sentence $S$ and an index $i$, we add it to the position $i$ of a paragraph $P$. We refer to the resultant text of this addition as $P\textprime$. As a result, 
\begin{equation*}
\label{eq:MR2t}
BE_t(P\textprime) = BE_t(P) \cup BE_t(S)
\end{equation*}
If the sentence is added before the start of the paragraph (i.e. $i = 0$), 
\begin{equation*}
BE_p(P\textprime)=
                  \{x \mid x\ \in BE_p(S)\} \cup \{x + length(S) \mid x \in BE_p(P)) \}
\end{equation*}
If the sentence is added to the end of the paragraph,
\begin{equation*}
BE_p(P\textprime)=
                  \{x \mid x\ \in BE_p(P)\} \cup \{x + length(P) \mid x \in BE_p(S) \}
\end{equation*}
If the sentence is added to the middle of the paragraph ($i$ is neither the start nor the end of the paragraph),
\begin{multline*}
\label{eq:MR2p3}
BE_p(P\textprime)=
                  \{x \mid x \in BE_p(P) \land x<i) \} \cup \\ \{x + length(S) \mid x \in BE_p(P) \land x>=i) \} \cup \\ \{x + i \mid x \in BE_p(S)) \} 
\end{multline*}
\subsubsection{MR3: Adding a paragraph to an article}
\begin{figure}[h]
\centering
\includegraphics[width=0.45\textwidth]{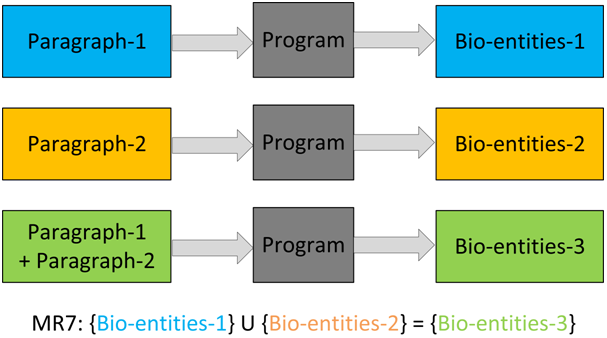}
\caption{\label{fig:examplemr} MT testing of LingPipe using MR3}
\end{figure}

Given a paragraph $P$ and an index $i$, we add this paragraph to the position $i$ of an article $A$. We refer to the resultant text of this addition as $A\textprime$. As a result, the following equations should be true.
\begin{equation*}
\label{eq:MR3t}
BE_t(A\textprime) = BE_t(A) \cup BE_t(P)
\end{equation*}
If the paragraph is added before the start of the article (i.e. $i = 0$), 
\begin{equation*}
BE_p(A\textprime)=
                  \{x \mid x\ \in BE_p(P)\} \cup \{x + length(P) \mid x \in BE_p(A) \}
\end{equation*}
If the paragraph is added to the end of the article,
\begin{equation*}
BE_p(A\textprime)=
                  \{x \mid x\ \in BE_p(A)\} \cup \{x + length(A) \mid x \in BE_p(P) \}
\end{equation*}
If the paragraph is added to the middle of the article (i.e. $i$ is neither the start nor the end of the article),
\begin{multline*}
BE_p(A\textprime)=
                  \{x \mid x \in BE_p(A) \land x<i) \} \cup \\ \{x + length(P) \mid x \in BE_p(A) \land x>=i) \} \cup \\ \{x + i \mid x \in BE_p(P)) \} 
\end{multline*}
A high-level depiction of MR3 is given in Figure \ref{fig:examplemr}.

\subsubsection{MR4: Adding a list of random words to another list of random words}
Given two list of random words $L_1$ and $L_2$, we append $L_2$ to $L_1$ and refer to the resultant list as $L\textprime$. In this MR, we make sure that $L_2$ is added to $L_1$ with a newline in-between to avoid combining the words at the interconnecting position. We should have the following relations:
\begin{equation*}
\label{eq:MR4t}
BE_t(L\textprime) = BE_t(L_1) \cup BE_t(L_2)
\end{equation*}
\begin{equation*}
\label{eq:MR4p}
BE_p(L\textprime) = \{x \mid x\ \in BE_p(L_1)\} \cup \{x+length(L_1) \mid  x \in BE_p(L_2) \}
\end{equation*}

\subsection{Deletion Relations}

\subsubsection{MR5: Removing a list of words from a sentence}
Given a sentence $S$, we remove a list of words, $L$, from $S$. Referring to the resultant text of this deletion as $S\textprime$, 
\begin{equation*}
\label{eq:MR5t}
BE_t(S\textprime) = BE_t(S) - BE_t(L)
\end{equation*}
\begin{equation*}
\label{eq:MR5p}
BE_p(S\textprime) = \{x - c \mid x \in BE_p(S) \land c \in \mathbb Z_{\ge 0} \}
\end{equation*}

\subsubsection{MR6: Removing a sentence from a paragraph}
Given a paragraph $P$ and an index $i$, we remove a sentence $S$, starting from position $i$, from $P$. We refer to the resultant text of this deletion as $P\textprime$. Therefore,
\begin{equation*}
\label{eq:MR6t}
BE_t(P\textprime) = BE_t(P) - BE_t(S)
\end{equation*}
If the first sentence of the paragraph is removed (i.e. $i = 0$), 
\begin{equation*}
BE_p(P\textprime)=
                  \{x - length(S) \mid x \in BE_p(P) \land x \notin BE_P(S)\}
\end{equation*}
If the last sentence of the paragraph is removed,
\begin{equation*}
BE_p(P\textprime)=
                  \{x \mid x \in BE_p(P) \land x \notin BE_P(S) \}
\end{equation*}
If a sentence is removed from the middle of the paragraph (i.e. $i$ is neither the start nor the end of the paragraph),
\begin{multline*}
BE_p(P\textprime)=
                  \{x \mid x \in BE_p(P) \land x<i) \} \cup \\ \{x - length(S) \mid x \in BE_p(P) \land x \notin BE_p(S) \land x>i) \} 
\end{multline*}
\subsubsection{MR7: Removing a paragraph from an article}
Given an article $A$, we remove a random paragraph $P$ from $A$. Referring to the resultant text of this deletion as $A\textprime$, 
\begin{equation*}
\label{eq:MR7t}
BE_t(A\textprime) = BE_t(A) - BE_t(P)
\end{equation*}
If the first paragraph of the article is removed (i.e. $i = 0$), 
\begin{equation*}
BE_p(A\textprime)=
                  \{x - length(P) \mid x \in BE_p(A) \land x \notin BE_P(P)\}
\end{equation*}
If we remove he last paragraph of the article is removed,
\begin{equation*}
BE_p(A\textprime)=
                  \{x \mid x \in BE_p(A) \land x \notin BE_P(P) \}
\end{equation*}
If a paragraph is removed from the middle of the article (i.e. $i$ is neither the start nor the end of the article),
\begin{multline*}
BE_p(A\textprime)=
                  \{x \mid x \in BE_p(A) \land x<i) \} \cup \\ \{x - length(P) \mid x \in BE_p(A) \land x \notin BE_p(P) \land x>i) \} 
\end{multline*}
\subsubsection{MR8: Removing some words from a list of random words}
Given a list of random words $L_1$, we remove half of these words from the end of file, and refer to the removed part and the remaining part as $L_2$ and $L\textprime$, respectively. We should have the following relations.
\begin{equation*}
\label{eq:MR8t}
BE_t(L\textprime) = BE_t(L_1) - BE_t(L_2)
\end{equation*}
\begin{equation*}
BE_p(L\textprime)=
                  \{x \mid x \in BE_p(L_1) \land x \notin BE_P(L_2) \}
\end{equation*}

\subsection{Shuffling Relations}
\subsubsection{MR9: Shuffling paragraphs of an article}
Given an article $A$, we shuffle all the paragraphs of $A$ and refer to the new resultant text as $A\textprime$. As a result, following equation must be true.
\begin{equation*}
\label{eq:MR9t}
BE_t(A\textprime) = BE_t(A)
\end{equation*}
With respect to this MR, the positions of bio-entities ($BE_p$s) can vary and there is no predefined relation between them.
\subsubsection{MR10: Shuffling a list of random words}
Given a list of random words $L$, we shuffle all words and create a new list of words $L\textprime$. Therefore, 
\begin{equation*}
\label{eq:MR10t}
BE_t(L\textprime) = BE_t(L) 
\end{equation*}
MR10 does not satisfy a predefined relation on $BE_p$s. Note that we feed these words to LingPipe as separate words.

\section{METHODOLOGY}
\label{sec:methodology}
\begin{figure}[h]
\centering
\includegraphics[width=0.45\textwidth]{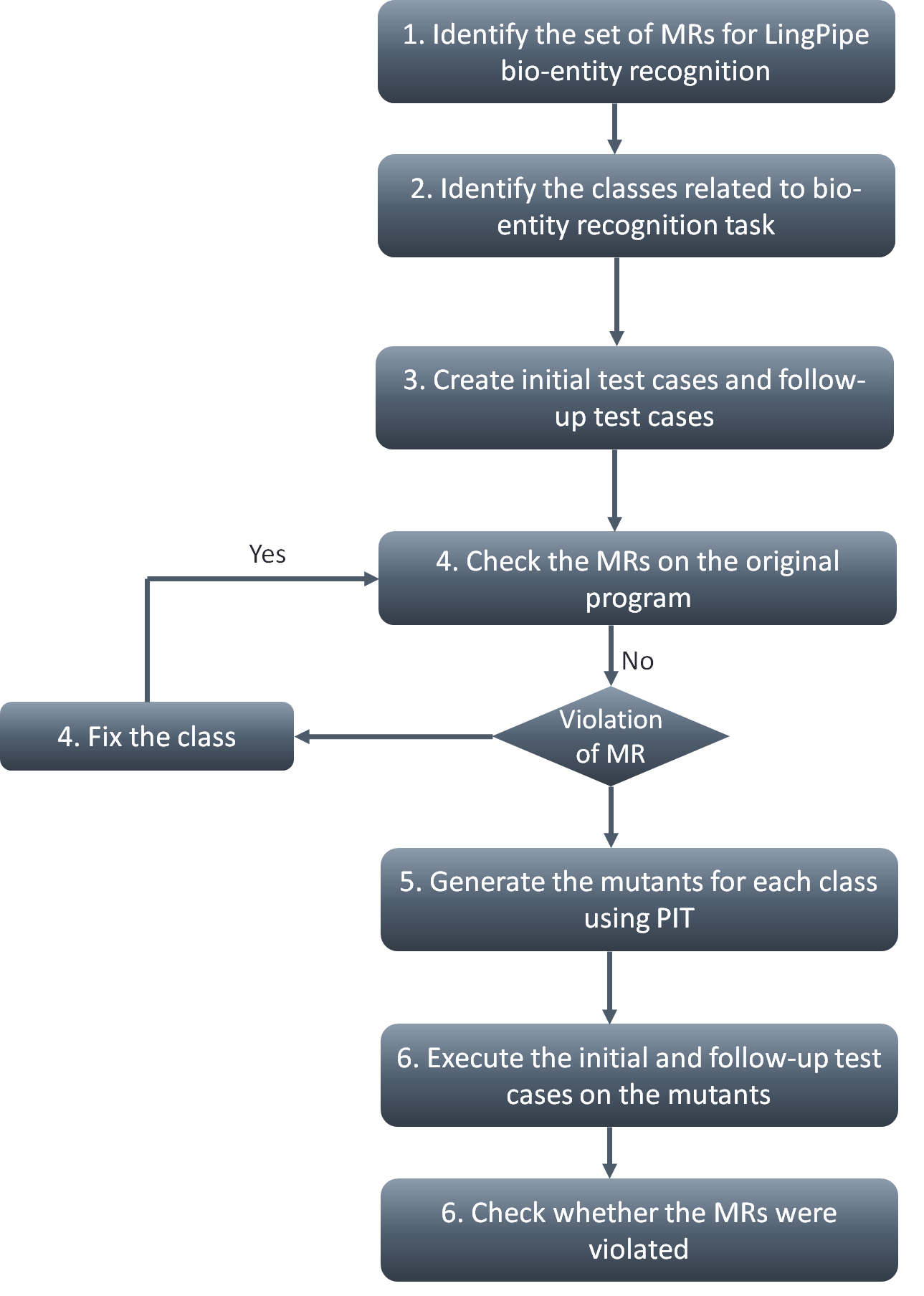}
\caption{\label{fig:flowchartlingpipe1}MT process for testing LingPipe}
\end{figure}

In this section, we describe our methodology for testing LingPipe 4.1.2 using MT. Figure~\ref{fig:flowchartlingpipe1} depicts the sequence of steps required to perform MT. Below is the sequence of steps involved.
\begin{enumerate}
\item Identify a set of MRs for testing the bio-entity recognition task in LingPipe. The MRs that we developed are described in section~\ref{section_mt}.

\item Develop source test cases and follow-up test cases for each of the MRs. Below, we describe how we created them. Note that all the articles, paragraphs, sentences, and words are extracted from a collection of biomedical articles available through the PubMed~\footnote{https://www.ncbi.nlm.nih.gov/pubmed} website.
\begin{enumerate}[I.]
\item MR1  (Adding a sentence to another sentence)
\begin{itemize}
\item Source test case: Two sentences extracted from a biomedical article (PMCID: 100320).
\item Follow-up test case: The sentence formed by appending one of them to the other sentence.
\end{itemize}
\item MR2  (Adding a sentence to a paragraph)
\begin{itemize}
\item Source test case: A paragraph and a sentence obtained from biomedical text (PMCID: 100320).
\item Follow-up test case: The sentence is added to the end of the paragraph forming an extended paragraph. 
\end{itemize}
\item MR3  (Adding a paragraph to an article)
\begin{itemize}
\item Source test case: A full-text biomedical article (PMCID: 100320) and a paragraph from another article (PMID: 28881451).
\item Follow-up test case: The paragraph is appended to the end of the article.
\end{itemize}
\item MR4  (Adding a list of random words to another list)
\begin{itemize}
\item Source test case: Two lists of five hundred random words extracted from 15 biomedical articles.
\item Follow-up test case: Append the first list to the second list to form a new list of thousand random words.
\end{itemize}
\item MR5  (Removing a list of random words from a sentence)
\begin{itemize}
\item Source test case: A sentence extracted from a biomedical article (PMCID: 100320) and a consecutive list of words from this sentence.
\item Follow-up test case: Remove the consecutive list of words from the sentence to create a truncated sentence.
\end{itemize}
\item MR6  (Removing a sentence from a paragraph)
\begin{itemize}
\item Source test case: A paragraph obtained from a biomedical article (PMCID: 100320) and a randomly selected sentence from that paragraph. 
\item Follow-up test case: A new truncated paragraph created by removing the randomly selected sentence.
\end{itemize}
\item MR7  (Removing a paragraph from an article)
\begin{itemize}
\item Source test case: A biomedical article and a randomly selected paragraph from that article (PMCID: 100320).
\item Follow-up test case: Remove the randomly selected paragraph from the article.
\end{itemize}
\item MR8  (Removing some words from a list of random words)
\begin{itemize}
\item Source test case: A list of one thousand random words extracted from 15 biomedical articles and the second half of this list that contains five hundred words.
\item Follow-up test case: First five hundred words of this list.
\end{itemize}
\item MR9  (Shuffling paragraphs of an article)
\begin{itemize}
\item Source test case: A biomedical article (PMCID: 100320).
\item Follow-up test case: A new article created by shuffling the paragraphs.
\end{itemize}
\item MR10  (Shuffling a list of random words)
\begin{itemize}
\item Source test case: A list of a thousand random words extracted from 15 biomedical articles.
\item Follow-up test case: a new list formed by shuffling all the above words.
\end{itemize}
\end{enumerate}

\item \label{step:classID}Identify set of classes related to the bio-entity recognition task. LingPipe 4.1.2 is written using Java and is a large software system consisting of 25 Java packages. Each of the packages contains a number of Java classes varying from three to 32 classes. We selected the classes related to bio-entity recognition task based on the class hierarchy structure. We also verified that these selected classes contain functionality related to bio-entity recognition task by examining execution traces.  

\item The source and follow-up test cases for each MR are executed on LingPipe. The MT process is put on hold if any of the MRs are violated. The errors in the program which caused the fault are identified and fixed. The execution of test cases is conducted on the rectified program. This process continues until all the MRs are satisfied (note: we did not find any violations of MRs on the LingPipe).

\item Mutants are generated for each of the classes identified in Step~\ref{step:classID}. Mutants are the faulty versions of the program under test and are created by introducing a single syntactic change in the source code. More information about the mutation generation process can be found in Section~\ref{sec:mutGen}. 
\item The source and follow-up test cases for each MR is executed on the set of mutants. The results of the source and follow-up test cases are checked to identify if the corresponding MR is violated. Violation of a MR is recognized if the corresponding mutant is \emph{killed}. 

\end{enumerate}

\section{Experimental Setup}
\label{sec:expsetup}
In this section we provide the details of the experimental setup, especially the research questions to be answered, the system under test (SUT), description of the Java classes that were tested and information regarding mutants generated for each class.
\subsection{Research Question}
\label{sec:researchquestion}
\begin{enumerate}
\item [\textbf{RQ1.1:}] How effectively does MT identify faults in the system under test? 
\item[\textbf{RQ1.2:}] Does MT identify more faults than unit test cases available in the system under test?
\item[\textbf{RQ2:}] Which MRs perform better in identifying faults in the system under test?
\end{enumerate}
\subsection{LingPipe-4.1.2}
LingPipe is a tool for processing text using computational linguistics~\cite{lingpipe}. LingPipe can help in performing tasks such as named entity recognition, sentiment analysis on news feeds and suggesting spelling corrections. We selected LingPipe as our case study since LingPipe is one of the most popular biomedical bio-entity recognition programs in the bioinformatics community. The tool also performs quick extraction of bio-entities from large paragraphs and articles which provides an ideal platform to consider LingPipe in our case study. 
\subsection{Java classes under test}
LingPipe provides a total of about 350 Java classes and each Java class performs functionality related to text mining. We determined that 15 classes are related to the bio-entity recognition task in particular. These classes were identified based on the dynamic sequence diagrams generated using the Jive tool \footnote{https://www.cse.buffalo.edu/jive/latest.html}. Two classes out of these 15 classes were selected for this study by examining the class hierarchy and execution trace. We conducted the mutation analysis using these two classes.
Functionality of each of the two selected classes are explained below:

\begin{enumerate}
\item \textbf{ChunkTagHandler} 
\newline
The ChunkTagHandler class sets the chunk handler to the specified bio-entity. This class deals with the arrays of tokens, tags, and white spaces. ChunkTagHandler converts the three arrays to a chunking and then passes the chunking to the chunk handler. The main functionality includes identifying bio-entities and dividing into chunks. The chunks are associated with a tag handler for extraction and representation~\cite{lingpipe}.
\item \textbf{IndoEuropeanTokenizer}
\newline
IndoEuropeanTokenizer returns a tokenized version of the required string. An array of tokens is constructed using the words, characters, and white spaces from the paragraphs. The generated tokens are grouped into chunks by the above mentioned ChunkTagHandler class~\cite{lingpipe}.
\end{enumerate}
\vspace{5 mm} %
\begin{table}
\centering
\begin{tabular}{|l|c|c|}
\hline
\ Java Class & \ \# LOC\\
\hline
ChunkTagHandler & 268\\
IndoEuropeanTokenizer & 310\\
\hline
\end{tabular}
\caption{ LOC (lines of code) for LingPipe classes}

\label{LingPipe classes}
\end{table}
\subsection{Mutant Generation}
\label{sec:mutGen}
Mutation testing is used in our experiments to determine the quality and effectiveness of the MRs to identify faults in the program. Mutants are generated for the two classes using the PIT tool \footnote{http://pitest.org/}. The faulty versions of the program are created using \emph{mutation operators}. Mutation operators apply changes to a statement in the program which creates fault in the program. Below, we describe the mutation operators used by PIT to generate the mutants:
\begin{enumerate}
\item \textbf{Conditional Operator:} The conditional boundary mutator replaces the relational operators $<$, $\leq$, $>$, $\geq$ with the alternate operator.
\item \textbf{Increment Operator:} The increment operator ++, \text{--} will replace increments with decrements and vice versa.
\item \textbf{Math Operator:} The math operators such as = -,*,\%, \&,$\ll$,$\gg$ are replaced by alternate operator in the list.
\item \textbf{Negate Conditional Operator:} The conditional operators such as ==, !=, $\leq$, $\geq$, $<$, $>$ will be replaced by alternate operator from the list.
\item \textbf{Return value mutator Operator:} The return types such as boolean, long, float double will be replaced by \textit{null}, \textit{zero}, \textit{1}, \textit{true} or \textit{false}. 
\end{enumerate}
Table ~\ref{table:mutant-java class} shows the total number of mutants generated, \emph{equivalent} mutants detected, mutants that generated exceptions for each class and the final set of mutants used for MT. The second column provides details of the total mutants generated for the two classes. The third column indicates the total mutants generating exceptions during the execution. Mutants causing exceptions such as the program crashing abruptly or leading to an infinite loop in execution of the program are removed from further consideration. Further, the mutants that produce the same outputs as the original program were manually inspected to filter out any mutants that were outside the scope of the bio-entity recognition task. The last column in Table~\ref{table:mutant-java class} presents the total number of mutants used in the experiment after filtering the mutants using the above criteria.



\begin{table*}[h]
\centering
\begin{tabular}{|l|c|c|c|c|c}
\hline
Java Class & \multicolumn{1}{p{3cm}|}{\centering Total \# of\\ Mutants generated} & \multicolumn{1}{p{3cm}|}{\centering \# of Mutants\\ giving exceptions} & \multicolumn{1}{p{3cm}|}{\centering \# of Mutants \\ with equal outputs} & \multicolumn{1}{p{3cm}|}{\centering \# Mutants tested \\ with MT}\\
\hline
ChunkTagHandler & 28 & 6 & 16 & 6\\
IndoEuropeanTokenizer & 64 & 18 & 15 & 31\\
\hline
\end{tabular}

\caption{\# Mutants for each Java class}
\label{table:mutant-java class}
\end{table*}
\section{Results}
\label{sec:results}
In thi section, we discuss our findings for each of the research questions described in section 5. 
\subsection{RQ1.1 \& RQ1.2: Effectiveness of MT}

The MT process, described in Section~\ref{sec:methodology}, killed 65\% (24 out of 37) of the total mutants for the two classes used for the evaluation. Based on the killing percentage, we can conclude that MT effectively detects faults in LingPipe.

\begin{table}[h]
\centering
\begin{tabular}{|l|c|c|c|}
\hline
 Java Class & \multicolumn{1}{p{2cm}|}{\centering Mutants killed\\ by unit testing} & \multicolumn{1}{p{2.3cm}|}{\centering Mutants killed\\ by MT approach}\\
\hline
ChunkTagHandler & not available & 5 (83\%)\\
IndoEuropeanTokenizer & not available & 19 (62\%)\\
\hline
\end{tabular}
\caption{\# Mutants killed by unit testing vs Metamorphic Testing approach}
\label{table:unit test case and MT comparison}
\end{table}
Table ~\ref{table:unit test case and MT comparison} shows the individual mutant killing rates for the ChunkTagHandler class and the IndoEuropeanTokenizer class. Neither of them has any unit test cases provided by the developers of Lingpipe. Thus, we could not compare the fault detection effectiveness of MT with the existing unit tests for those two classes. 

\subsection{RQ2: Effectiveness of individual metamorphic relations for fault detection}
Figure~\ref{fig:killingrate} and Table~\ref{table:NumberOfKilledMutants} show the mutant killing rates of the individual MRs for the the two classes used in the experiment. For the ChunkTagHandler class, MR9 (shuffling the paragraphs of an article) and MR7 (removing a paragraph from an article) has killed the most mutants (67\% each). The least performing MR is MR10 (shuffling the words) within a list of random words and it only killed 17\%  of the mutants. 

In the IndoEuropean class, MR9 (shuffling the paragraphs) of an article performed best by killing 55\% mutants, followed by the second best performing MR7 (removing a paragraph) from an article killing 48\% mutants. Interestingly, the least performing MR10 ( shuffling a set of random words) did not kill any mutant. 

When comparing the overall mutant killing rates of the MRs for both the classes, MR9 (shuffling the paragraphs) of an article performed best by providing 56\% overall killing rate. The second best performing MR7 (removing a paragraph) from an article provides an overall mutant killing rate of 51\%. The least performing MR10 (shuffling the words) within a list of random words provided an overall 2\% mutant killing rate.

\begin{table}[h]
\centering
\begin{tabular}{|c|c|c|c|}
\hline
 MR\# &  \multicolumn{1}{p{3cm}|}{\centering ChunkTagHandler \\ (out of 6)} &  \multicolumn{1}{p{3cm}|}{\centering IndoEuropeanTokenizer \\ (out of 31)}\\
\hline
MR1 & 3 & 14 \\
MR2 & 3 & 11 \\
MR3 & 3 & 7 \\
MR4 & 3 & 12 \\
MR5 & 3 & 12 \\
MR6 & 3 & 11 \\
MR7 & 4 & 15 \\
MR8 & 3 & 12 \\
MR9 & 4 & 17 \\
MR10 & 1 & 0 \\
\hline
\end{tabular}
\caption{\# Mutants killed for each class by the individual MRs.}
\label{table:NumberOfKilledMutants}
\end{table}

\begin{center}
\begin {figure*}
\begin{adjustbox}{width=\textwidth}
\begin{tikzpicture}
\begin{axis}[
axis lines*=left,
	ybar,
	symbolic x coords={MR1,MR2,MR3,MR4,MR5,MR6,MR7,MR8,MR9,MR10},
    xtick={MR1,MR2,MR3,MR4,MR5,MR6,MR7,MR8,MR9,MR10},
	ylabel=\% of mutants killed,
    xlabel=List of MRs,
    xlabel near ticks,
    title=Killing Rate,
    bar width=0.01cm,
    width=.4\textwidth,
    width=14cm,height=5cm,
    ymin=0,ymax=100,
    bar width=0.4cm,
    xticklabel style={rotate=0},
    yticklabel style = {font=\small},
    xticklabel style = {font=\small},
    nodes near coords={\pgfmathprintnumber\pgfplotspointmeta\%},
    nodes near coords align={vertical},
    nodes near coords style={scale=0.75},
    legend style={at={(0.5,-0.3),font=\small},
	anchor=north,legend columns=-1},
]
\addplot 
	coordinates {(MR1,50) (MR2,50) (MR3,50) (MR4,50) (MR5,50) (MR6,50) (MR7,67) (MR8,50) (MR9,67) (MR10,17) };
    \addplot 
	coordinates {(MR1,45) (MR2,35) (MR3,23) (MR4,39) (MR5,39) (MR6,35) (MR7,48) (MR8,39) (MR9,55) (MR10,0)};
\legend{ChunkTagHandler,IndoEuropeanTokenizer}
\end{axis}
\end{tikzpicture}
\end{adjustbox}
\captionof{figure}{Killing rate of MRs for Java classes}
\label{fig:killingrate}
\end{figure*}
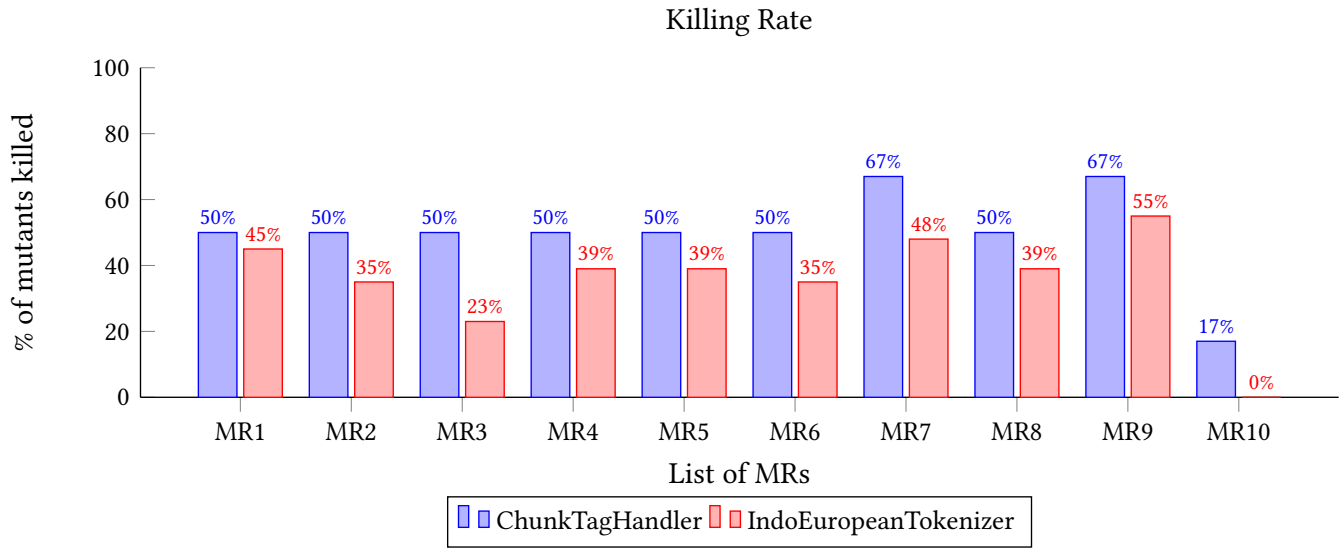
\end{center}

\section{Discussion}
\label{sec:discussion}
In this study, we developed 10 novel MRs for bio-entity recognition and evaluated their effectiveness for conducting MT using a widely used bio-entity recognition tool, LingPipe.

\textbf{Feasibility of MT approach for bio-entity recognition:} 
Our experiment results show that MT killed 83\% of the mutants for the ChunkTagHandler class and 61\% of the mutants for the IndoEuropeanTokenizer class. Thus, we could detect most of the mutants using the developed MRs showing that MT is an effective method for testing bio-entity recognition tools. 

Both the classes we used in the experiments provided important functionality for bio-entity recognition task in LingPipe. But unfortunately, neither of these classes had any unit tests provided by the developers. This could be due to the oracle problem associated with the bio-entity recognition task. Therefore, MT provides an effective approach for quality assurance of these important classes especially in the absence of unit tests. 


\textbf{Performance of individual MRs in fault detection:}
In the IndoEuropeanTokenizer class, the best performing MRs, MR7 (removing a paragraph from an article) and MR9 (shuffling the paragraphs of an article) covered more statements compared to the other MRs. Also these two MRs showed different execution paths in the IndoEuropeanTokenizer class for the source and the follow-up test cases. In the ChunkTagHandler class, these two MRs showed different execution paths for the source and follow-up test cases but the number of statements covered by these MRs were not different from the other MRs.

The experiment results also show that MR9 and MR7 killed the same set of mutants except one mutant that was only killed by MR7. This indicates the possibility that the MR9 is redundant. 

Overall, MR10 (shuffling a set of random words) had the lowest mutant killing rate indicating that MR10 might be a less effective MR compared to the rest of the MRs. But, interestingly, MR10 killed a mutant of the ChunkTagHandler class which could not be killed by any of the other MRs. 

\section{Related Work}
\label{sec:relatedworks}
Over the past years, there has been an enormous growth in the quantity and variety of biological data. With this growth, researchers have developed a lot of programs in bioinformatics. But, most of time and effort is spent on developing complex statistical methods, while there is a lack of effort in systematically testing these programs that is essential for the quality assurance~\cite{chen2009innovative}. 

Chen et al.~\cite{chen2009innovative} employed MT on two open-source bioinformatics programs. The first program GNLab ~\footnote{http://en.bio-soft.net/other/gnlab.html} is a tool for large-scale analysis and simulation of gene regulatory networks. The second program SeqMap~\footnote{http://www-personal.umich.edu/~jianghui/seqmap/} deals with mapping a short sequence that reads with a reference genome. 

They introduced new MRs for these programs and showed that MT is beneficial in detecting faults as well as its relative ease of use. Pullum and Ozmen~\cite{pullum2012early} employed MT for testing epidemiological models. They used the model parameters based on ordinary differential equation and agent based model and the expected results which are obtained from executing the model with MR-transformed parameter values. Ramanathan et al.~\cite{ramanathan2012verification} also utilized MT to build a work flow of compartments of susceptible, infectious, and recovered epidemiological models. They showed that MT can be useful where the mathematical models may fail. Anders et al.~\cite{Upulee} examined the effectiveness of a pseudo-oracle and MT to detect subtle faults in bioinformatics program. The results showed that MT performed better than pesudo-oracles for detecting subtle faults. To the best of our knowledge, our study is the first time that MT is applied for the quality assurance of bio-entity recognition.

\section{Conclusions and Future Work}
\label{sec:conclusions}
MT is a technique to test programs which do not have a test oracles. Bio-entity recognition tools face the oracle problem since it produces a very large number of bio-entities for a given text corpus making it hard to validate. In this study, we examined the effectiveness of MT for the quality assurance of bio-entity recognition. First, we proposed 10 novel MRs for validating bio-entity recognition tools in general. Then we applied these MRs for testing a popular bio-entity recognition tool, LingPipe. Our results show that MRs that we developed are effective in identifying faults in the tool. 

In our future work, we plan to increase the mutant killing percentages by creating source test cases based on an effective test case generation mechanism as an alternative to the used random test cases. We also plan to combine MRs so that we can increase their effectiveness while reducing the cost associated with executing multiple MRs. We will also extend our evaluation to include other classes associated with bio-entity recognition in LingPipe.

\bibliographystyle{ACM-Reference-Format}
\bibliography{myRef}

\end{document}